\documentstyle[12pt,prd,aps,epsfig,preprint]{revtex}
\begin{document}
\draft \tightenlines
\date{\today}

\title{A Stochastic Mean-Field Approach For Nuclear Dynamics}
\author{Sakir Ayik}
\address{Physics Department, Tennessee Technological University, Cookeville, TN 38505, USA}

\maketitle

\begin{abstract}

We propose a microscopic stochastic approach to improve description of nuclear dynamics beyond the mean-field approximation at low energies. It is shown that, for small amplitude fluctuations, the proposed model gives a result for the dispersion of a one-body observable that is identical to the result obtained previously through a variational approach. Furthermore, by projecting the proposed stochastic mean-field evolution on a collective path, a generalized Langevin equation is derived for collective variable, which incorporate one-body dissipation and one-body fluctuation mechanism in accordance with quantal fluctuation-dissipation relation.

\end{abstract}


\section{Introduction}

In the mean-field description of a many-body
system, the time-dependent wave function  is assumed
to be a Slater determinant constructed with $A$ number of
time-dependent single-particle wave functions $\Phi_j(\vec{r},t)$.
The single-particle wave-functions are determined by the
Time-Dependent Hartree-Fock equations (TDHF) with proper initial
conditions \cite{Ring,Goeke,Davis},
\begin{eqnarray}
i\hbar\frac{\partial}{\partial t}\Phi_j(\vec{r},t)
=h(\rho) \Phi_j(\vec{r},t)
\end{eqnarray}
where $h(\rho)$ denotes the self-consistent mean-field Hamiltonian.
Rather than the single-particle wave functions,
in many situations, it is more appropriate to express the mean-field
approximation in terms of the single-particle density matrix
$\rho(\vec{r},\vec{r}^{\prime},t)$, which is defined as
\begin{eqnarray}
\rho(\vec{r},\vec{r}^\prime,t)=\sum_{j} \Phi_j^{*}(\vec{r},t)  n_j
\Phi_j(\vec{r}^\prime,t)
\end{eqnarray}
where $n_j$ denotes occupation factors of the single-particle states. In standard TDHF, there are $A$ number of occupied states for which the occupation factors are one, $n_j=1$, and zero for unoccupied states. If the initial state is at a finite temperature $T$, occupation factors are determined by the Fermi-Dirac distribution, $n_j=1/[exp(\epsilon_j-\mu)/T+1]$. The single-particle density matrix in the mean-field approximation evolves according to the transport equation,
\begin{eqnarray}
i\hbar \frac {\partial }{\partial t}\rho(t)=[h(\rho), \rho(t)]
\end{eqnarray}
The expectation value $ Q(t)$ of a one-body observable $\hat{Q}$ and its dispersion $\sigma^{2}_{Q}(t)$ are calculated according to,
\begin{eqnarray}
Q(t)=\sum_{j} <\Phi_j(t)|\hat{Q}|\Phi_j(t)> n_{j}
\end{eqnarray}
and
\begin{eqnarray}
\sigma^{2}_{Q}(t)=\sum_{kj} |<\Phi_k(t)|\hat{Q}|\Phi_j(t)>|^{2} n_{k}(1-n_{j})
\end{eqnarray}
The mean-field approximation includes, so called, the one-body dissipation mechanism and therefore it provides a good approximation for the average evolution of the collective motion at sufficiently low energies around  $10 MeV$ per nucleon, at which two-body dissipation and fluctuation mechanism do not have an important influence on dynamics. In the mean-field approximation, while the single-particle motion is treated in the quantal framework, the collective motion is treated nearly in the classical approximation. Therefore, the TDHF provides a good description of the average evolution of collective motion, however it severely restricts the fluctuations of the collective motion.

Much effort has been given to improve the mean-field description by incorporating two-body dissipation and fluctuation mechanisms \cite{AyikGregoire,Randrup,Abe}. The resultant stochastic transport theory provides a suitable framework for dissipation and fluctuation dynamics of nuclear collisions at intermediate energies. However, the two-body dissipation and fluctuation mechanism does not play an important role at low energies. Fundamental question is that how to improve the mean-field dynamics by incorporating one-body fluctuation mechanism (i.e, mean-field fluctuations) associated with the one-body dissipation in a microscopic level?  In this work, we address this question. We restrict our treatment at low energies at which mean-field evolution and the one-body dissipation mechanism provide a good approximation and propose a simple stochastic mean-field approach for describing fluctuation dynamics. In section 2, we give a brief description of the stochastic approach. In section 3, we illustrate that the stochastic approach gives rise to the result, which is identical a previous formula derive using a variational approach. In section 4, by projecting stochastic mean-field evolution on a collective path, we derive a generalized Langevin equation for the collective variable. Conclusions are given in section 5.

\section{Stochastic TDHF Equation}

In the mean-field framework only source for stochasticity can arise from the initial correlations. In the standard approach, the TDHF eq.(3)  provides a deterministic evolution for the single-particle density matrix, starting from a well defined initial state. On the other hand, as a result of the correlations, the initial state can not have a well-defined single determinantal form, but rather it must be a superposition of determinantal wave functions. In the mean-field approximation, even though correlations are not propagated, initial correlations can be incorporated by considering, not a single initial state but a distribution of initial Slater determinants. It is well-known that, such initial correlations can be simulated in a stochastic description \cite{Mori}. In order to develop a stochastic description, we need to determine sufficient number of unoccupied single-particle states in addition to the occupied ones, and write the single-particle density matrix in the form,
\begin{eqnarray}
\rho^{\lambda}(\vec{r},\vec{r}^\prime,t)=\sum_{ij}
\Phi_i^{*}(\vec{r},t;\lambda)  ~\rho^{\lambda}_{ij}(t_0)~\Phi_j(\vec{r}^\prime,t;\lambda)
\end{eqnarray}
Here $\rho^{\lambda}_{ij}(t_0)$ are the time-independent elements of density matrix which is determined by the initial conditions. The main assumption of the approach is that each matrix element is a Gaussian random number specified by a mean value $\overline \rho^{\lambda}_{ij}(t_0)=\delta_{ij} n_j $, and a variance
\begin{eqnarray}
\overline { \delta \rho^{\lambda}_{ij}(t_0)~
\delta\rho^{\lambda}_{j^\prime i^\prime}}(t_0) =\frac{1}{2}
\delta_{ii^\prime} \delta_{jj^\prime}[n_i(1-n_j)+ n_j(1-n_i)]
\end{eqnarray}
In these expressions, $\lambda$ represents the event label and $\delta \rho^{\lambda}_{ij}(t_0)$ represents the fluctuating elements of initial density matrix, $\rho^{\lambda}_{ij}(t_0)=\delta_{ij} n_j+\delta \rho^{\lambda}_{ij}(t_0)$. At zero temperature, the occupation numbers are one or zero, and at finite temperature $n_j$ are determined by the Fermi-Dirac distribution. In each event, different from the standard TDHF,
the time-dependent single-particle wave functions are determined by its own self-consistent mean-field according to,
\begin{eqnarray}
i\hbar\frac{\partial}{\partial t}\Phi_j(\vec{r},t;\lambda)
=h(\rho^{\lambda}) \Phi_j(\vec{r},t;\lambda)
\end{eqnarray}
where $h(\rho^{\lambda})$ denotes the self-consistent mean-field Hamiltonian in the event $\lambda$.
Similar to eq. (3), we can express the stochastic mean-field evolution in terms of the single-particle density matrix as,
\begin{eqnarray}
i\hbar \frac {\partial}{\partial t} \rho^{\lambda}(t) =[h(\rho^{\lambda}), \rho^{\lambda}(t)]
\end{eqnarray}
As we illustrate in the following sections, in the stochastic mean-field approach, one-body dissipation and fluctuation mechanism is incorporated into dynamics in a manner consistent with the quantum fluctuation-dissipation relation, in a microscopic level.

\section{Fluctuations of One-Body Observables}

In the proposed stochastic approach, employing eq.(9) with random initial conditions, we generate an ensemble of different events. Each event evolves according to its own self-consistent mean-field potential.
In this approach, not only the mean value of an observable represented by a
one-body operator $\hat{Q}$, but we can calculate the probability distribution of the observable. The expectation value of an observable $\hat{Q}$ in an event is determined as,
\begin{eqnarray}
Q^{\lambda}(t)=\sum_{ij} <\Phi_i(t;\lambda)|\hat{Q}|\Phi_j(t;\lambda)> \rho^{\lambda}_{ji}(t_0).
\end{eqnarray}
We note that the influence of the initial fluctuations appears in
both the elements of the density matrix and also matrix elements of
the observable. Even if the magnitude of the initial fluctuations is
small, the mean-field evolution can enhance the fluctuations,
and hence events can
substantially deviate from one another. The mean value of the
observable is determined by taking an average over the ensemble
generated by numerical simulations, $Q(t)=\overline{Q^{\lambda}}(t)$.
In a similar manner, the
variance of the observable is calculated according to,
\begin{eqnarray}
\sigma_Q^2(t)= \overline{(Q^{\lambda}(t)-Q(t))^2}
\end{eqnarray}

Here, we consider that the fluctuations are small, $\rho^{\lambda}(t)= \rho(t)+ \delta \rho^{\lambda}(t)$, where $\delta \rho^{\lambda}(t)$ shows the small amplitude fluctuations of the density matrix around the average evolution $\rho(t)$. In this case, in expression (10) the correlations between the matrix elements of the observable and the initial density matrix can be neglected and the mean value of the observable is calculated in the usual manner and
given by the standard TDHF result,
\begin{eqnarray}
Q(t)=\sum_{j} <\Phi_j(t)|\hat{Q}|\Phi_j(t)> n_j
\end{eqnarray}
In this expression, the single-particle wave functions are determined according to the standard TDHF equations with the standard mean-field Hamiltonian.

In order calculate the variance of a one-body observable, we need to determine the fluctuations of the density matrix
$\delta \rho^{\lambda}(t)$. Small fluctuations of the density matrix is determined by the time-dependent RPA equation, which is obtained by linearizing the stochastic TDHF equation (9) around the average evolution,
\begin{eqnarray}
i\hbar \frac {\partial}{\partial t} \delta \rho^{\lambda}(t) =
[\delta h^{\lambda}, \rho(t)]+ [h(\rho), \delta \rho^{\lambda}(t)].
\end{eqnarray}
Here, $\delta h^{\lambda}(t)=(\partial h/\partial\rho)\cdot \delta
\rho(t)$ denotes the fluctuating part of the mean-field Hamiltonian.
Employing a complete set of stationary single-particle representation, we can express the matrix elements of the time-dependent RPA equations as,
\begin{eqnarray}
i\hbar \frac {\partial}{\partial t} \delta \rho^{\lambda}_{ij}(t) =
\sum_{kl} R_{ij,kl}(t)~ \delta \rho^{\lambda}_{lk}(t)
\end{eqnarray}
where, $<\Phi_i|\delta \rho^{\lambda}(t)|\Phi_j>=
\delta \rho_{ij}^{\lambda}(t)$ denotes the elements of density matrix and the matrix $R(t)$ is given by
\begin{eqnarray}
R_{ij,kl}(t)&=& <\Phi_i|h(\rho)|\Phi_l> \delta_{kj}-
\delta_{il}<\Phi_k|h(\rho)|\Phi_j>+  \nonumber \\
&& \sum_{n} \left[ <\Phi_i\Phi_k|\partial h/\partial
\rho|\Phi_n\Phi_l> \rho_{nj}(t)-
\rho_{in}(t)<\Phi_n\Phi_k|\partial h/\partial
\rho|\Phi_j\Phi_l> \right]
\end{eqnarray}
Using super-space notation, eq.(14) can be written in a compact form as,
\begin{eqnarray}
i\hbar \frac{\partial}{\partial t}<ij|\delta \rho^{\lambda}(t)>=
<ij|R(t)|\delta \rho^{\lambda}(t)>
\end{eqnarray}
or
\begin{eqnarray}
i\hbar \frac{\partial}{\partial t}|\delta \rho^{\lambda}(t)>=
R(t)|\delta \rho^{\lambda}(t)>
\end{eqnarray}
where $|\delta \rho(t)>$ acts like a vector with double indices and $R(t)$ is a matrix in this super-space. Because of linear form, formal solution of this equation is given by,
\begin{eqnarray}
|\delta \rho^{\lambda}(t)>=
\exp\left[-\frac{i}{\hbar}\int_{t_0}^{t}R(s)ds\right]
|\delta \rho^{\lambda}(t_0)>
\end{eqnarray}
The fluctuating part of an observable is calculated according to
\begin{eqnarray}
\delta Q^{\lambda}(t_1)=
<Q|\exp\left[-\frac{i}{\hbar}\int_{t_0}^{t_1}R(s)ds\right]
|\delta \rho^{\lambda}(t_0)>
\end{eqnarray}
where $t_1$ represent the final time at which the observation is made and $t_0$ is the initial time. It maybe more convenient to define a time-dependent
one-body operator $B(t)$ according to
\begin{eqnarray}
<B(t)|=< Q~exp\left[-\frac{i}{\hbar}\int_{t}^{t_1}R(s)ds\right]|
\end{eqnarray}
It is easy to show that time evolution of the Heisenberg operator $B(t)$ is determined by the dual of the time-dependent RPA according to
\begin{eqnarray}
i\hbar \frac {\partial}{\partial t} B(t) = [h(\rho), B(t)]+
tr\left(\frac{\partial h}{\partial \rho}\right)\cdot [B(t),\rho]
\end{eqnarray}
The solution is determined by backward evolution this equation with the boundary condition $B(t_1)=Q$. In an event $\lambda$, the expectation value of the observable can be expressed as,
\begin{eqnarray}
\delta Q^{\lambda}(t_1)=<B(t_0)|\delta \rho^{\lambda}(t_0)>=
\sum_{ij}<\Phi_i(t_0)|B(t_0)|\Phi_j(t_0)>\delta \rho^{\lambda}_{ji}(t_0)
\end{eqnarray}
Then, the variance of the observable is calculated as,
\begin{eqnarray}
\sigma^2_Q(t_1)&=& \sum <\Phi_i(t_0)|B(t_0)|\Phi_j(t_0)>
<\Phi_{j^\prime}(t_0)|B(t_0)|\Phi_{i^\prime}(t_0)>
\overline{\delta\rho_{ji}^{\lambda}(t_0)~ \delta\rho_{i^{\prime}j^{\prime}}^{\lambda}(t_0)}\nonumber \\
&=& \sum |<\Phi_i(t_0)|B(t_0)|\Phi_j(t_0)>|^2n_i(1-n_j)
\end{eqnarray}
where the last equality is obtained using the formula (7). This result is identical with the formula derived in a previous work employing variational approach \cite{Balian}.

\section{Projection on Collective Path}

In order to illustrate the fact that the stochastic TDHF equation describes dynamics of fluctuations in accordance with the one-body dissipation mechanism, we give another example in this section. We consider that the collective motion is slow and maybe describe by a few relevant collective variables. For example, in induced fission dynamics relevant collective variables
maybe taken as the relative distance of fragments, mass-asymmetry and neck parameter. Here, we consider a single collective variable $q(t)$, and introduce the quasi-static single-particle representation,
\begin{eqnarray}
h(q) \Psi_j(\vec{r};q)=\epsilon_j(q) \Psi_j(\vec{r};q)
\end{eqnarray}
where $h(q)=h[\rho(q)]$ denotes the mean-field Hamiltonian, in which the time dependence of the local density  $\rho(\vec{r},\vec{r};q)$ is parameterized in terms of the collective variable in a suitable manner. We expand the
single-particle density in terms of the quasi-particle representation,
\begin{eqnarray}
\rho(\vec{r}, \vec{r}^{\prime};q)=
\sum_{kl} \Psi_k^{*}(\vec{r};q) \rho_{kl}(t)\Psi_l(\vec{r}^{\prime};q).
\end{eqnarray}
Both the elements of density matrix $\rho_{kl}(t)$ and collective variable $q(t)$ are fluctuating quantities. In this section for clarity of notation, we ignore the event label $\lambda$ on these quantities. We determine the matrix elements of density in the lowest order perturbation theory in dynamical coupling $ <\Psi_k|\partial \Psi_l/\partial q>\dot{q}(t)$. It is preferable that the wave functions are close to diabatic structure. Since in the diabatic representation, dynamical coupling is expected to be small, hence, it can be treated in the weak-coupling approximation. Diabatic
single-particle representation can approximately be constructed by ignoring small symmetry breaking terms in the mean-field potential \cite{Noerenberg,AyikNoerenberg}.

In order to determine the temporal evolution of collective variable, we use the total energy conservation,
\begin{eqnarray}
E=\sum_{lk} <\Psi_k(q)|T|\Psi_l(q)> \rho_{lk}(t)+
\frac{1}{2} \sum_{ijlk} \rho_{ji}(t) <\Psi_k(q)\Psi_i(q)|V|\Psi_l(q)\Psi_j(q)>\rho_{lk}(t)
\end{eqnarray}
In the many-body Hamiltonian, for simplicity we take an effective two-body interaction potential energy $V$. The total energy depends on time implicitly via collective variable $q(t)$ and explicitly via matrix elements $\rho_{lk}(t)$. Energy conservation requires,
\begin{eqnarray}
\frac{dE}{dt}=\dot{q}\frac{\partial E}{\partial q}+
\frac{\partial E}{\partial t}=0
\end{eqnarray}
In this expression $-\partial E/\partial q$ represents a dynamical force acting on the collective variable. The force depends on time, and it evolves from an initial diabatic form accompanied with deformation of Fermi surface towards an adiabatic limit associated with the adiabatic potential energy. The second term represents the rate of change of the energy due to explicit time dependence,
\begin{eqnarray}
\frac{\partial E}{\partial t}= \sum_{lk} <\Psi_k(q)|h(\rho)|\Psi_l(q)>
\frac {\partial}{\partial t} \rho_{lk}(t)= \sum_{k} \epsilon_k(q)
\frac{\partial}{\partial t} \rho_{k}(t).
\end{eqnarray}
Here, $\rho_{k}(t)=\rho_{kk}(t)$ represents the occupation factors of the quasi-static single-particle states. It is possible to derive a master equation for the occupation factors by substituting the expansion (25) into eq.(9) to give
\begin{eqnarray}
\frac{\partial }{\partial t} \rho_{kl}= -\frac{i}{\hbar} \epsilon_{kl}~\rho_{kl} +\sum_{j} \left[\rho_{kj}<\Psi_j|\partial \Psi_l/\partial q>-
<\Psi_k|\partial \Psi_j/\partial q>\rho_{jl}\right]\dot{q}(t)
\end{eqnarray}
where $\epsilon_{kl}=\epsilon_k(q)-\epsilon_l(q)$. This can be
converted into an integral equation as,
\begin{eqnarray}
\rho_{kl}(t)&=& \int_{t_0}^{t}dt_1 \sum_{j} \
G_{kl}(t,t_1)\left[\rho_{kj}(t_1)<\Psi_j|\partial \Psi_l/\partial
q>_{t_1}\right.
- \nonumber \\
&& \left. <\Psi_k|\partial \Psi_j/\partial q>_{t_1}\rho_{jl}(t_1)
\right]\dot{q}(t_1)  + G_{kl}(t,t_0)\rho_{kl}(t_0)
\end{eqnarray}
where $G_{kl}(t,t_1)=
exp [-\frac{i}{\hbar}\int_{t_1}^{t}\epsilon_{kl}(q)ds]$ is the mean-field propagator and $\rho_{kl}(t_0)$ denotes the initial value of the matrix element.
Substituting this expression in the right hand side of eq.(29) and keeping only diagonal elements of density matrix, we obtain a master equation for the occupation factors,
\begin{eqnarray}
\frac{ \partial}{\partial t} \rho_{k}(t)&=& \int_{t_0}^{t}dt_1 \sum_{j}
\left [ G_{jk}(t,t_1) <\Psi_k|\partial \Psi_j/\partial q>_{t} \dot{q}(t)
<\Psi_j|\partial \Psi_k/\partial q>_{t_1}\dot{q}(t_1)+c.c \right](\rho_k-\rho_j)- \nonumber\\
&& \sum_{j} \left[G_{jk}(t,t_0)<\Psi_k|\partial \Psi_j/\partial q>_{t}
\dot {q}(t)\rho_{jk}(t_0)+c.c.\right].
\end{eqnarray}
This is a stochastic master equation for the occupation factors. Therefore, it determines not only the mean value of the occupation factors, but also their distribution functions. The first term on the right hand side is the one-body collision term determined by the dynamical coupling, while the last term represents its stochastic part. The stochasticity arises from the initial correlations $\rho_{jk}(t_0)$, as described in section 2. As a result, the stochastic part of the collision term has a Gaussian distribution with zero mean and a variance specified in terms of the variance of the initial matrix elements according to eq. (7).

Using the master equation for the occupation factors, the rate of change of total energy due to explicit time dependence can be expressed as,
\begin{eqnarray}
\frac{ \partial E}{\partial t}&=& \int_{t_0}^{t}dt_1 \sum_{jk}
G_{jk}(t,t_1)\left[ \epsilon_{kj} <\Psi_k|\partial \Psi_j/\partial
q>_{t} \dot{q}(t)
<\Psi_j|\partial \Psi_k/\partial q>_{t_1}\dot{q}(t_1)+c.c \right](\rho_k-\rho_j)- \nonumber\\
&& \sum_{jk}[G_{jk}(t,t_0) \epsilon_{kj}<\Psi_k|\partial \Psi_j/\partial q>_{t}
\dot {q}(t)\rho_{jk}(t_0)+c.c.].
\end{eqnarray}
The magnitude of the coupling matrix elements, $<\Psi_k|\partial \Psi_j/\partial
q>$, should decrease as a function of energy difference, mainly as a result of the mismatch of overlap of wave functions. The energy dependence of the average behavior of square of coupling matrix element may be described by a Gaussian or Lorentzian form factor.
In weak-coupling limit, that we consider here, decay time of the memory kernel is determined by the correlation time of the coupling matrix elements defined as
$\tau_c=\hbar/\Delta$, where $\Delta$ is the energy range of the form factor of the coupling matrix elements. During time interval smaller than the correlation time, $t-t_1 < \tau_c $, we neglect the variation of single-particle energies as a function of collective variable and approximate the mean-field propagator by
$G_{kl}(t,t_1)= exp [-\frac{i}{\hbar}(t-t_1)\epsilon_{kl}]$.
Furthermore, for slow collective motion it is possible to carry out an
expansion in powers of $t-t_1$, as it was done in the linear response
treatment of ref.\cite{Hofmann}. Here, we do not carry out such an expansion, but consider the collective motion in a parabolic potential well . In
this case, the memory effect can be taken approximately into account by
incorporating harmonic propagation of collective motion during
short time intervals according to,
\begin{eqnarray}
\dot{q}(t_1)=\dot{q}(t)cos\Omega(t_1-t)+
\frac {1}{\Omega}\ddot{q}(t)sin\Omega(t_1-t)
\end{eqnarray}
where $\Omega$ is the curvature of the harmonic well. Incorporating the result into eq.(32), and factoring out $\dot{q}(t)$ from each term, we deduce a generalized Langevin equation of motion for the collective variable \cite{Takigawa,Ayik},
\begin{eqnarray}
M \ddot{q} + \frac{1}{2} \frac {dM}{dq} \dot{q}^2+ \frac{\partial
E}{\partial q} = -\gamma~ \dot{q} +\xi(t) \end{eqnarray}
where $M$, $\gamma$ and $\xi$ denotes the inertia, the friction
coefficient and the stochastic force, respectively. These quantities
are given by,
\begin{eqnarray}
M =2\hbar^2 \sum_{jk} |<\Psi_k|\partial \Psi_j/\partial q>|^2~
\frac {\epsilon_{jk}}{(\epsilon_{jk})^2+(\hbar \Omega)^2}~\rho_k
\end{eqnarray}
\begin{eqnarray}
\gamma(\Omega) =\sum_{jk} |<\Psi_k|\frac{\partial h}{\partial q}|\Psi_j>|^2
\frac {1}{\Omega}
\left[\frac {\eta}{(\epsilon_{jk}-\hbar \Omega)^2+\eta^2}-
\frac {\eta}{(\epsilon_{jk}+\hbar \Omega)^2+\eta^2} \right]
\rho_k (1-\rho_j)
\end{eqnarray}
and
\begin{eqnarray}
\xi(t)= \sum_{jk} G_{jk}(t,t_0) <\Psi_k|\frac{\partial h}{\partial
q}|\Psi_j> \rho_{jk}(t_0)
\end{eqnarray}
where $\epsilon_{jk}= \epsilon_j-\epsilon_k$. A similar treatment can be carried out for a harmonic potential barrier \cite{Ayik}. In obtaining the expressions for the inertia and the friction
coefficient, we let the upper limit of time integration in eq.(32) go
to infinity, $t-t_0\rightarrow \infty$. We notice that the principle
part of the integration determines the inertia and real transitions
described by the delta-function part of the propagator determines the friction
coefficient. We note that the inertial force term in eq.(34)
for the collective variable arises from the $q(t_1)$ dependence on
the coupling matrix element in eq.(32). This dependence is
included by expanding $<\Psi_j|\partial \Psi_k/\partial q>_{t_1}$
around $q(t)$ in the lowest order in $q(t_1)-q(t)$ and using
harmonic propagation for $q(t_1)$ according to eq.(33). Also, a
contribution to friction arises from this term, which vanishes at
zero frequency limit. Here, we ignore this contribution to the
friction. The expression of the inertia is different from the
adiabatic mass formula, which is known as Cranking formula \cite{Ring}. Due to
factor $\epsilon_{jk}$ in nominator, unlike the adiabatic mass
formula, large contributions to inertia near quasi-crossings vanishes, and hence  rapid variations of the adiabatic inertia in the vicinity of quasi-crossing of
levels do not show up in the expression (35).

As a result of stochastic properties of the initial correlations, stochastic force has a Gaussian distribution with zero mean $\overline{\xi}(t)=0$ and a second moment determined by the autocorrelation function,
\begin{eqnarray}
\overline{\xi(t) \xi(t_1)}= \sum_{jk} G_{jk}(t,t_1)
|<\Psi_j|\frac{\partial h}{\partial q}|\Psi_k>|^2\frac{1}{2}
\left[\rho_j(1-\rho_k)+ \rho_k(1-\rho_j)\right].
\end{eqnarray}
We introduce the Fourier transform of the autocorrelation function of the stochastic force. After a straightforward manipulation, it is easy to show that
the autocorrelation function is related to frequency spectrum of the friction coefficient according to
\begin{eqnarray}
\overline{\xi(t) \xi(t_1)}=
\int_{-\infty}^{+\infty}\frac{d\omega}{2\pi} e^{-i
\omega(t-t_1)}~\hbar \omega~ coth \frac{\hbar
\omega}{2T}~\gamma(\omega)
\end{eqnarray}
where $\gamma(\omega)$ is given by eq.(36) in which the collective frequency $\Omega$ is replaced by the running frequency $\omega$. This result represents the quantum fluctuation-dissipation relation associated with the one-body dissipation mechanism and it naturally emerges from the stochastic approach presented in this work \cite{Gardiner,Weiss}. In the expression of the friction coefficient involves a form factor arising from energy dependence of the matrix elements. As discussed above, the form factor can be taken as a Gaussian or Lorentzian specified by the correlation energy range $\Delta$. The correlation energy acts as a cut-off for the integral over the frequency in eq.(39). At high temperature limit, $T>>\Delta$, since
$coth(\hbar\omega/2T)=2T/\hbar\omega$, the expression (39) reduces the  classical result,
\begin{eqnarray}
\overline{\xi(t) \xi(t_1)}\approx 2T~\gamma_{0}~\delta(t-t_1)
\end{eqnarray}
where $\gamma_0=\gamma(0)$ is the zero frequency limit of the friction coefficient. In the zero temperature limit, on the other hand, $coth(\hbar\omega/2T)\approx 1$, and the autocorrelation function,
\begin{eqnarray}
\overline{\xi(t) \xi(t_1)}=
\int_{-\infty}^{+\infty}\frac{d\omega}{2\pi} e^{-i
\omega(t-t_1)}~\hbar\omega~\gamma(\omega)
\end{eqnarray}
purely associated with quantum zero point fluctuations.

\section{Conclusion}

In this work, we propose a microscopic stochastic approach to improve description of nuclear dynamics beyond the mean-field approximation at low energies. In order to demonstrate scientific validity of the approach, we present two illustrations. In the first case, we show that, for small amplitude fluctuations, the proposed stochastic model gives a result for the dispersion of a one-body observable that is identical to previous result obtained through a variational approach. In the second case, by projecting the stochastic
mean-field evolution on a collective path, we derive a generalized Langevin equation for collective variable, which incorporate one-body dissipation and one-body fluctuation mechanism in accordance with quantal
fluctuation-dissipation relation. We believe that such illustrations provide a strong support that the proposed model provides a consistent microscopic description for dynamics of density fluctuations in nuclear collisions at low energies. In fact, such a stochastic approach provides a useful tool not only for nuclear dynamics but also other fields, such electron dynamics in atomic and condense matter physics. Numerical simulations of the unconstraint 3-D TDHF equations have been carried out with powerful computational tools. In the proposed stochastic description, we need to calculate a sufficient number of events by numerical simulations. The numerical effort for such mean-field calculations is not very extensive and can easily be accomplished with the present day computational power.

\section{Acknowledments}

The author gratefully acknowledges Middle East Technical University
and GANIL for warm hospitality extended to him during his visits.
This work is supported in part by the US DOE Grant No.
DE-FG05-89ER40530. Also, the author thanks to D. Boilley, A, Gokalp,
D. Lacroix, B. Yilmaz and O. Yilmaz for fruitful discussions.


\end{document}